\documentstyle[12pt,epsfig]{article}

\makeatletter
\def\La{\langle}
\def\Ra{\rangle}
\def\ra{{\rm{a}}}
\def\re{{\rm{e}}}
\def\rf{{\rm{f}}}
\def\ru{{\rm{u}}}
\def\rd{{\rm{d}}}
\def\rs{{\rm{s}}}
\def\rc{{\rm{c}}}
\def\rb{{\rm{b}}}
\def\rQ{{\rm{Q}}}
\def\rQbar{\bar{\rm{Q}}}

\def\ccdot{\hbox{\kern-.1em$\cdot$\kern-.1em}}

\newlength{\abstwidth}
\begin{document}

\pagestyle{empty}

\begin{flushright}
CERN--TH/97--294  \\
% Leiden \\
hep-ph/9710462
%hep-ph/9708261
\end{flushright}
 
\vspace{\fill}
 
\begin{center}
{\LARGE\bf Meson--photon transition form factors and}\\[3mm]
{\LARGE\bf resonance cross-sections in $\re^+\re^-$ collisions}\\[10mm]
{\Large G.\ A.\ Schuler$^{a,b}$, F.\ A.\ Berends$^{a,c}$, 
R.\ van Gulik$^c$} \\[2ex]
{\it ${}^a$ Theory Division, CERN,
%} \\[1mm]
%{\it 
CH-1211 Geneva 23, Switzerland}
%\\[1mm]
%{ E-mail: {\tt Gerhard.Schuler@cern.ch}}
\\[2ex]
%{\large and} \\[2ex]
%{\Large Frits Berends and Robert Gulik}\\[3mm]
{\it ${}^c$ Instituut--Lorentz,
%}\\[1mm]
%{\it 
University of Leiden,
%}\\[1mm]
%{\it 
The Netherlands}
%\\[1mm]
%{E-mail: {\tt gulik@rulil0.LeidenUniv.nl}}
\end{center}
 
\vspace{\fill}
 
\begin{center}
{\bf Abstract}\\[2ex]
\begin{minipage}{\abstwidth}
Meson--photon--photon transition form factors 
for $S$-, $P$-, and $D$-wave states are calculated, 
the meson being treated as a non-relativistic heavy-quark--antiquark pair. 
The full dependence on both photon virtualities is included. 
Cross-section formulas for charge-conjugation even mesons with 
$J^P = 0^-$, $0^+$, $1^+$, $2^+$, and $2^-$ in electron--positron 
collisions are presented and numerical results for LEP energies are given. 
In particular, we find two-photon event rates for $\chi_{\rc 1}$, 
$\eta_{\rc}(2S)$, and $\eta_{\rb}(1S)$ within reach of LEP. 
 
With minor modifications to incorporate $SU(3)$-flavour breaking 
we estimate rates for $18$ light mesons as well, based on the observation 
that their two-photon decay widths agree remarkably well 
with measured data. Finally we point out that $\re^+\re^-$ cross sections 
for $1^+$ states do not vanish at low $Q^2$, the Landau--Yang suppression 
factors of the two-photon cross sections being compensated by the photon 
propagators.
\end{minipage}
\end{center}

\vspace{\fill}
\noindent
\rule{60mm}{0.4mm}

\vspace{1mm} \noindent
%${}^{a}$ Talk presented at the XIth Workshop on Photon--Photon Collisions, 
%Photon '97, Egmond aan Zee, The Netherlands.\\[1mm]
${}^b$ Heisenberg Fellow.

\vspace{10mm}\noindent
CERN--TH/97--294\\
October 1997

\clearpage
\pagestyle{plain}
\setcounter{page}{1}

%{\it Stored in ~/public/exclusive/res.tex.}

\section{Introduction}
In this paper we discuss resonance production in two-photon fusion 
in high-energy e\-lec\-tron--positron ($\re^+ \re^-$) collisions. 
The main focus of our study is the dependence of the cross section 
on the photon virtualities $Q_i^2$, which we take fully into account. 
While there exist several papers on the $Q^2$ dependence of the 
(single) form factor that governs the production of 
pseudo-scalar mesons \cite{BL,ref:PionFF,Kroll97,Reviews},
much less is known about the (in general several) form factors 
associated with any other charge-conjugation even ($C=+1$) meson 
\cite{KWZ,KV,Cahn} (for experimental data, see 
\cite{Poppe,Photon97}). 
%the scalar mesons \cite{},
%the tensor mesons \cite{},
%the pseudo-tensor mesons \cite{},
%the axial-vector mesons \cite{Cahn}. 

Here we present analytical 
results for the form factors and cross sections of $C=+1$ mesons up to 
and including $D$-wave states. We compare these results 
with existing calculations where available. 
Moreover, we give numerical results for $30$ mesons 
at typical LEP energies, for various representative experimental setups. 
All formulas are implemented in the Monte Carlo event generator 
GALUGA \cite{GALUGA}, and hence cross sections and distributions for 
any kind of $\re^+\re^-$ environment can be easily obtained.

Let us now briefly discuss the theoretical framework in which
our results are derived.
Our calculation starts from the limit of heavy quarks, in which
case a meson can be considered a non-relativistic bound state 
of a heavy quark ($\rQ$) and a heavy antiquark ($\rQbar$).  
Corrections from both the motion of the heavy quarks within the meson 
and higher-Fock-state components are small. The theoretical description 
of production and decay of heavy quarkonia is based on the 
NRQCD factorization framework \cite{BBL}, where relativistic corrections and 
higher-Fock-state contributions are suppressed by powers of $v$, 
the relative velocity of the quarks in the meson. 

The velocity $v$ is reasonably small for charmonia and bottomonia, 
$\La v^2 \Ra \sim 0.3$ and $0.1$, respectively, so that reliable  
results can be expected for heavy quarkonia. 
On the contrary, the application of this approach to light mesons 
(consisting of $\ru$, $\rd$, and $\rs$ quarks) 
can certainly not be derived from QCD and has to be considered as a model. 
However, we shall see that, with only minor modifications, the 
two-photon widths of essentially all measured light mesons 
agree remarkably well with data. Hence we proceed to present 
form factors and cross sections for the light mesons as well. 

The dominant contribution to two-photon production of 
mesons arises from the (exclusive) short-distance process where 
a $Q\bar{Q}$ pair is produced with quantum numbers equal to those
of the asymptotic meson, i.e.\ where the meson is produced 
in its dominant Fock state. For a meson with
total angular momentum $J$, parity $P$ and charge-conjugation $C$
this is a colour-singlet $Q\bar{Q}$ pair with total spin $S$ ($=0,1$) 
and orbital angular momentum $L$ ($=0,1,2,\ldots$ or $S$, $P$, $D$, etc.)
such that ${\bf J} = {\bf L} + {\bf S}$, $P=(-1)^{L+1}$ and $C=(-1)^{L+S}$. 
Specifically, for the five $C=+1$ mesons of $J^P=0^-$, $J^+$, $2^-$
considered in this paper, the dominant Fock state is a $Q\bar{Q}$ state with
${}^{2S+1}L_J = {}^1S_0$, ${}^3P_J$, ${}^1D_2$, respectively 
(in the spectroscopic notation and $J=0,1,2$). 

Four of the above mesons are 
produced in two-photon fusion without any short-distance suppression. 
Hence no enhanced $O(\alpha_s)$ and $O(v^2)$ corrections can occur and 
estimates based on the lowest-order $O(\alpha^2\, \alpha_s^0)$ calculation 
should be reliable. 
The situation could be different for the $1^+$ state. Its 
$O(\alpha^2)$ cross section vanishes when both photons are real and 
is therefore suppressed by $\sim \left< Q^2\right> /M^2$ 
(details will be given below). Hence, if the reaction is not totally 
exclusive, other processes can be important. 
The $Q^2$ suppression of the exclusive process 
$\gamma\gamma \rightarrow R$ then competes with the 
suppression by extra powers of $\alpha_s(m)$ and/or $v^2$ 
of inclusive resonance production, which, however, are finite at zero $Q_i^2$.

To leading order in $\alpha_s$ and/or $v$, 
three different mechanisms for the inclusive resonance 
production can occur for real photons. 
First, the short-distance production of a $\rQ\rQbar_1({}^3P_1)$. 
Here and in the following a subscript $1$ ($8$) indicates a 
colour-singlet (colour-octet) $\rQ\rQbar$ pair. 
This production proceeds via $\gamma\gamma \rightarrow 
\rQ\rQbar_1({}^3P_1)+gg$ 
or $\rQ\rQbar_1({}^3P_1)+\rQ\rQbar$ and is suppressed by 
$\alpha_s^2\, v^0$. 
Secondly, there is the $\alpha_s^0\, v^4$-suppressed mechanism 
where a $\rQ\rQbar_1({}^3P_{0,2})$ pair, produced at short distances, 
turns into the $1^+$ state via a double $E1$ transition. 
Both contributions are small with respect to a third mechanism that is 
suppressed merely by a power of $\alpha_s(m)$: a 
$\rQ\rQbar_8({}^3S_1)$ pair is produced, which then turns into 
a $\rQ\rQbar_1({}^3P_1)$ via a single $E1$ transition. The power 
of $v^2$ of the $E1$ transition is compensated by the fact that 
$S$-wave production is favoured against $P$-wave production by $1/v^2$. 
Whether or not any of the above-mentioned processes
is important depends on the experimental setup: all these 
mechanisms are characterized by the presence of at least one more pion. 
Here we assume that the  experimentally selected events are truly single
resonance states  
so that these contributions are absent.

In the following sections we shall discuss helicity amplitudes and
widths for the decay of resonances (section 2), form factors for the
production (section 3) and finally results for the production
cross sections.

\section{Two-photon decays}
Consider now the decay of a $C=+1$ resonance into two photons, 
$  R(P) \rightarrow \gamma^\star(k_1) + \gamma^\star(k_2)$, 
and define $  W = \sqrt{P^2}$, 
$K_i = \sqrt{k_i^2}$, $\nu = k_1 \cdot k_2$, and 
$X = \nu^2 - K_1^2\, K_2^2$. While $W$ equals the resonance mass $M$ 
for the decay, the two variables may differ for the crossed reaction,
when one wants to take into account the Breit--Wigner formula in the
production.
We calculate the decay amplitudes using standard 
techniques\footnote{We modify the projection operators of \cite{KKS} 
to conform to the convention diag$(g_{\mu\nu}) = (+,---)$ and to include 
colour, and extend the rules to $D$-wave states.}.
We prefer to calculate helicity amplitudes rather than invariant 
amplitudes, since the former are more convenient to 
implement (after crossing) for the $\re^+\re^-$ cross section 
respecting the full $Q^2$ dependence. The amplitudes are
given in the resonance rest system. As spin direction for the
resonance we take the photon momentum.

Our results for the independent helicity decay amplitudes
$A(\lambda_1,\lambda_2)$ 
are listed in Table~\ref{tab:decayamplitudes} in units 
of $c_l = \sqrt{3/M}\, e_Q^2\, 16\, \pi\, \alpha\, R^{(l)}_{nl}(0)\, 
Y_{l0}(0,0) / D^{l+1}$ where $Y_{lm}(\theta,\phi)$ are the spherical 
harmonics, $D = W^2/4 - m^2 - \nu$. In NRQCD the resonance mass is twice
the quark mass, $M \approx 2\, m$. Here, 
$R^{(l)}_{nl}(0)$ is the $l$-th derivative of the radial wave function 
$R_{nl}(r) = \psi_{nlm}({\bf r}) / Y_{lm}(\theta,\phi)$ 
of the bound state at $r=0$. 
\begin{table}[tbp]
\begin{center}
\begin{tabular}{|c|c|c|c|c|c|}
\hline
$J^P$ & $A(+-)$ &  $A(++)$ &  $A(+0)$ &  $A(00)$ & $\Gamma_{\gamma\gamma}$
\\ \hline
$0^-$ & $0$     &  $\sqrt{X}$ & $0$ & $0$ & $4$
\\ \hline
$0^+$ & $0$ & $\frac{2\, (X+\nu\, W^2)}{\sqrt{3}\, W}$ & $0$
            & $\frac{-2}{\sqrt{3}}\, K_1\, K_2\, W$ & $144$
\\ \hline
$1^+$ & $0$ & $\frac{-\sqrt{2}\, \nu}{W}\, (K_1^2 - K_2^2)$ 
  & $\sqrt{2}\, K_2\, (\nu - K_1^2)$ & $0$  & $32$
\\ \hline
$2^+$ & $2\, W\, \nu$ & $\sqrt{\frac{2}{3}}\, 
  \frac{\nu\, \left( K_1^2 + K_2^2 \right) + 2\, K_1^2\, K_2^2 }{W}$ 
  & $\sqrt{2}\, K_2\, (\nu + K_1^2)$
  & $2\,  \sqrt{\frac{2}{3}}\,W\, K_1\, K_2$  & $192/5$
\\ \hline
$2^-$ & $0$ & $4\,  \frac{X^{3/2}}{M^2}$ 
  & $0$ & $0$      & $64$
\\ \hline
\end{tabular}
\caption[]{Decay amplitudes $A(\lambda_1,\lambda_2)$ in units of 
$c_l$ and two-photon decay width 
(reduced with $\tilde{\Gamma}_{\gamma\gamma}$ for $1^+$)
in units of $d_l$ (see text for the
definition of $c_l$, $d_l$). 
%$c_l = \sqrt{3}\, e_Q^2\, 16\, \pi\, \alpha\, R^{l}_{nl}(0)\, 
%Y_{l0}(0,0) / D^{l+1}$ where $Y_{lm}(\theta,\phi)$ are the spherical 
%harmonics and $D = W^2/4 - m^2 - \nu$.
\label{tab:decayamplitudes}}
\end{center}
\end{table}
The photon helicities can take on the values $\lambda_i = \pm 1,0$. 
The remaining helicity amplitudes can be obtained 
using the relations
\begin{eqnarray}
  A(\lambda_1,\lambda_2) & = &   \eta_R\, A(-\lambda_1,-\lambda_2) 
\label{parityrel}
\\
  A(\lambda_1,\lambda_2) & = &   (-1)^J\,
  \left.  A(\lambda_2,\lambda_1)\right|_{K_1 \leftrightarrow K_2}
\ , 
\label{Boserel}
\end{eqnarray}
where $\eta_R = 1$ ($-1$) for mesons of the ``normal'' 
(``abnormal'') $J^P$ series $J^P=0^+,1^-,2^+,\ldots$ 
($J^P=0^-, 1^+, 2^-, \ldots$). 
%The invariant amplitudes for the $P$-wave states obtained in intermediate 
%steps of the calculation agree with \cite{KKS}.

Our amplitudes are normalized such that the two-photon decay width 
is given by
\begin{equation}
\Gamma^{\gamma\gamma}[J^P] =  \frac{1}{2\, J+1}\, 
  \frac{1}{32\, \pi\, M}\, \sum_{\lambda_1,\lambda_2 = \pm 1} 
%         \int \frac{\rd \Omega}{4\, \pi}\,
 \left|  A(\lambda_1,\lambda_2) \right|^2
\ .
\end{equation}
The width formulas are also displayed in 
Table~\ref{tab:decayamplitudes} and agree with the literature, see 
e.g.\ \cite{Novikov}. Here we have defined 
$d_l = 3\, e_Q^4\, \alpha^2\, |R_{nl}^{(l)}(0)|^2 / M^{2(l+1)}$. 
In the case of the $1^+$ meson the entry defines the reduced 
width\footnote{Our $\tilde{\Gamma}_{\gamma\gamma}$ coincides with 
that of \cite{TPC} and is one half that of \cite{Cahn}.} 
$\tilde{\Gamma}_{\gamma\gamma}$. This is the transverse--transverse
two-photon width divided by a factor $[(K_1^2-K_2^2)/(2\, \nu)]^2$, 
which shows that  $\Gamma_{\gamma\gamma}[1^P]$ is zero, 
in agreement with the Landau--Yang theorem. 

Before presenting numerical results for the two-photon widths, 
we have to discuss the input parameters, namely the wave functions 
and the squared charge factor $e_Q^2$ appearing in $c_l$ and $d_l$. 
Obviously, the electric charge of the quark is $e_Q = +2/3$ ($-1/3$) 
for charmonia (bottomonia). Moreover, 
the wave functions for the heavy quarkonia can quite reliably be 
calculated by solving the Schr\"odinger equation with a phenomenological 
inter-quark potential. We take values from a recent 
potential-model calculation \cite{EQ}. The results are valid
in the large-mass limit, where the non-relativistic
expansion makes sense.

For the light mesons we have to assume the constituent quark model
to be still a good approximation. We start from a linear 
potential $\propto \lambda\, r$, in which case
$\left| R_{nS}(0) \right|^2 = 2\, \mu\, \lambda$ (independent of the radial 
quantum number $n$), 
$\left| R'_{1P}(0) \right|^2 = 0.268\, 
\left(2\, \mu\, \lambda \right)^{5/3}$, 
$\left| R''_{1D}(0) \right|^2 = 0.151\, 
\left(2\, \mu\, \lambda \right)^{7/3}$, 
where $\mu$ is the reduced mass. 
Using canonical values for the string tension $\lambda$ 
and constituent-quark masses, we
take $2\, \mu\, \lambda = 0.74\,$GeV$^3$.
%$\lambda = 0.3\, $GeV$^{2}$. 
%and constituent masses for the light quarks of a few hundred MeV
% of$m_u = m_d = 0.333\,$GeV 
%we get the following values
In order to incorporate $SU(3)$ breaking we multiply the above squared wave
functions by $r_M^{2 l + 1}$, $r_M = M/ \mu_0$, 
%\begin{equation}
%r_M = (\frac{M} {\mu_0})^{2l+1}
%\label{scale}
%\end{equation}
where $M$ is the meson mass and $\mu_0$ a hadronic scale of about 1 GeV. 
We thus use the following values  for $R^{(l)}_{nl}(0)$:
\begin{equation}
\begin{array}{llll}
%\hline
 & {\rm Light}\,\,\,{\rm mesons}\qquad & {\rm Charmonia}\qquad
 & {\rm Bottomonia}
\\ 
%\hline
\left| R_{1S}(0) \right|^2 / {\rm GeV}^3 \qquad & 0.074\, r_M & 0.81 & 6.5
\\ 
%\hline
\left| R'_{1P}(0) \right|^2 / {\rm GeV}^5 \qquad  & 3.5\times 10^{-3}\, r_M^3 
& 0.075 & 1.4
\\ 
%\hline
\left| R''_{1D}(0) \right|^2 / {\rm GeV}^7 \qquad 
 & 0.35\times 10^{-3}\, r_M^5
 & 0.015 & 0.64
\\ 
\left| R_{2S}(0) \right|^2 / {\rm GeV}^3 \qquad  & 0.074\, r_M & 0.53 & 3.2
\\ 
%\hline
\end{array}
\label{wavevalues}
\end{equation}  
For the light mesons we also have to replace $e_Q^2$ 
by the effective squared charge  
$\La e_q^2 \Ra$, which depends on the mixing angle $\theta$ characterizing 
the breaking of the $SU(3)$-flavour symmetry: 
%SU(3)$ breaking effects will be large, given the large spread 
%between the light meson masses, the $\ru$, $\rd$, $\rs$ masses, 
%and between the quark and meson masses. In our calculation we have 
%assumed a common quark mass $m \approx M/2$ for all light mesons. 
%The results will strongly depend on the way to implement mass breaking. 
%
%Our implemention of $SU(3)$ breaking is based on the following reasoning. 
%Consider first the $P$-wave states $J^+$ that can decay into two photons 
%with relative orbital angular momentum zero. Then it is natural to assume 
%that it is the helicity amplitudes that obey the relations of 
%$SU(3)$ mixing. 
%This amounts to replacing $e_Q^2$ in $A$ by
\begin{eqnarray}
%  \left< e_Q^2 \right> =
 \La e_q^2 \Ra = 
   & \frac{1}{3\, \sqrt{2} }
   & \pi, \ra_0, \ra_1, \ra_2, \pi_2
\nonumber\\
%  \left< e_Q^2 \right> =
   & \frac{1}{3\, \sqrt{6} } \, 
         \left( \cos\theta - 2\, \sqrt{2}\, \sin\theta \right) \qquad
   & \eta, \rf_0, \rf'_1, \rf'_2, \eta_D
\nonumber\\
%  \left< e_Q^2 \right> =
   & \frac{2}{3\, \sqrt{3} } \, 
         \left( \cos\theta + \frac{1}{2\, \sqrt{2}}\, \sin\theta \right)
  \qquad
   & \eta', \rf'_0, \rf_1, \rf_2, \eta'_D
\ .
\label{mesoncharge}
\end{eqnarray}
%On the other hand, 
%
%the P-wave helicity amplitudes $A$ behave as 
%$A \sim R'_P(0)/\sqrt{M^3}$. In order to make these 
%(meson-)mass independent, $R'_P(0)$ has to scale as $M^{3/2}$. 
%This prescription is implemented by multiplying the 
%two-photon decay widths by a factor $(M/\mu_0)^3$ where $\mu_0$ 
%is a hadronic scale of order $1\,$GeV. 

The incorporation of $SU(3)$ breaking outlined above 
can be refined by including the effect
of the centrifugal barrier. Parity allows the orbital
angular momentum of the photons to be 0 for the $J^+$ states, 
but requires at least 1 for the cases of $0^-$ and $2^-$. 
In our calculation this suppression shows up\footnote{
The additional factor $X/M^2$ visible in Table~\ref{tab:decayamplitudes} 
for the ${}^1D_2$ state arises from $k_1^\mu\, k_2^\nu\, 
\epsilon_{\mu\nu}(2,J_z)$ and is not counted as a threshold factor.}
as the factor $\epsilon_{\alpha\beta\mu\nu}\, k_1^\alpha\, k_2^\beta 
\, \epsilon_1^\mu\, \epsilon_2^\nu  \propto\, \sqrt{X}$ 
and therefore implies an additional factor $r_M^2$ ($r_M^4$) 
for the $0^-$ and $2^-$ helicity amplitudes (decay widths).
%Hence we demand $A[0^-,2^-]/M^2$ to obey the $SU(3)$ relations rather 
%than $A$ itself. Since the amplitudes behave as 
%$A[0^-] \sim R_{S}(0)/\sqrt{M}$ and $A[2^-] \sim R''_{D}(0)/\sqrt{M^5}$, 
%we have need $R_S$ and $R''_D$ to scale as 
%$R_S(0) \sim M^{5/2}$, $R''_D(0) \sim M^{9/2}$. 
Effectively this changes the power of $r_M$ in (\ref{wavevalues}) 
into $n(J^P)$, where
%In total end up with decay-width expressions for the light mesons
%which are obtained by the substitution
%\begin{equation}
%  e_Q^2 \Longrightarrow \La e_q^2 \Ra\, 
%\left(\frac{M}{\mu_0}\right)^{n(J^P)/2}
%\label{prescription}
%\end{equation}
%in $c_l$ and $d_l$ where 
$n(J^P) = 5,3,9$ for $J^P = 0^-, J^+, 2^-$. 
Hence the two-photon widths scale with the meson mass as $M^3$ for 
$0^-$ and $2^-$ states, and as $1/M$ for $J^+$ states. 
This can be compared with the ``conventional'' approach,
where also the $J^+$ mesons are assumed to scale as $M^3$ \cite{conv,GI}. 

\begin{table}
\[
  \begin{array}{|c|c|c|c|c|c|} \hline
    i \backslash j & I=1 & I=0 & (I=0)'     & c \bar{c} & b \bar{b} \\ \hline
0^- & \pi^0 & \eta & \eta' & \eta_{\rc}(1S) & \eta_{\rb}(1S)\\ 
  1\, ^1S_0 & \frac{7.74}{10^{3}} (\frac{7.74}{10^{3}})
          & 0.41 (0.46) 
          & 6.1 (4.2) 
          & 7.8 (7.5) 
          & 0.46 (\mbox{---}) 
\\ \hline
0^+ & \ra_0 & \rf_0 & \rf_0' & \chi_{\rc 0} & \chi_{\rb 0}\\ 
    ^3P_0 & 5.1 (> 0.24)        
          & 0.72 (0.56)
          & 10.4 (5.4) 
          & 2.5 (4.0)
          & 0.043 (\mbox{---}) 
\\ \hline
1^+ & \ra_1 & \rf_1 & \rf_1' & \chi_{\rc 1} & \chi_{\rb 1}\\ 
    ^3P_1 & 0.90 (\mbox{---}) 
          & 2.5 (2.4) 
          & 0.10  (\mbox{---}) 
          & 0.50   (\mbox{---}) 
          & 0.92/10^2 (\mbox{---}) 
\\ \hline
2^+ & \ra_2 & \rf_2 & \rf_2' & \chi_{\rc 2} & \chi_{\rb 2}\\ 
    ^3P_2 & 1.0 (1.0)
          & 3.0 (2.4)  
          & 0.12 (0.097) 
          & 0.28 (0.37)
          & 0.74/10^2 (\mbox{---})
\\ \hline
2^- & \pi_2 & \eta_D & \eta'_D & \eta_{\rc D} & \eta_{\rb D}\\ 
    ^1D_2 & 1.3 (1.35)
          & 0.43 (\mbox{---})   
          & 5.0 (\mbox{---})  
          & 0.95/10^2 (\mbox{---})
          & 0.74/10^4(\mbox{---}) 
\\ \hline
0^- & \pi(2S) & \eta(2S) & \eta'(2S) & \eta_{\rc}(2S)
      & \eta_{\rb}(2S)\\ 
   2\,  ^1S_0 & 6.9 (\mbox{---})   
          & 2.3 (\mbox{---})   
          & 23.0 (\mbox{---})  
          & 3.5 (\mbox{---})
          & 0.20(\mbox{---}) 
\\ \hline
  \end{array}
\]
\caption{The $\gamma \gamma$ widths of the resonances in keV 
(reduced width $\tilde\Gamma_{\gamma\gamma}$ in the case of ${}^3P_1$). 
Central values of the experimental measurements (in parentheses) 
from PDG \cite{PDG}, except for the ${}^3P_1$, which is from 
TPC/$2\gamma$ \cite{TPC}.
\label{tab:ggWidth}}
\end{table}
Numerical results for the two-photon widths of the $n=1$ ($n$ the radial 
quantum number) mesons with $J^P=0^-$, $J^+$, and $2^-$ are given 
in Table~\ref{tab:ggWidth} for charmonia and bottomonia as well as for 
the light mesons with isospin $I=0,1$. The masses of the mesons 
are taken from the PDG, where known, and the others from a potential-model
calculation \cite{GI}. 
The latter are the $\eta_b(9400)$, and the $D$-wave states 
$\eta_D(1680)$, $\eta'_D(1890)$, $\eta_{\rc D}(3840)$, and
$\eta_{\rb D}(10150)$ (particle masses in parentheses). 
We have boldly taken $\ra_0(983.5)$, $\rf_0(980)$, and $\rf_0(1370)$ 
as the lowest-lying isoscalar despite their questionable status. 
Values for the  first radial excitation are given as well, where we took 
$\pi[2S](1300)$, $\eta[2S](1295)$,  $\eta'[2S](1400)$,  
$\eta_{\rc}[2S](3594)$, $\eta_{\rb}[2S](9980)$. 

The only free parameters are the scale $\mu_0$ and the mixing angles. 
We do not try to fit these. Rather we adjust $\mu_0$ 
by the $\pi^0$ decay width to $\mu_0 = 0.96\,$GeV. 
We determine the mixing angles of the pseudo-scalars and the tensor mesons 
from the quadratic mass formula $\theta[0^-] = -11.5^\circ$, 
$\theta[2^+] = 32^\circ$. Lacking further information, 
we simply take $\theta = 32^\circ$ for the other $P$-wave states as well 
and $\theta=0$ for the $D$-wave and $n=2$ $S$-wave states. 
A look at Table~\ref{tab:ggWidth} shows that we find surprisingly 
good agreement with the measured decay widths for practically 
all measured mesons\footnote{For an (incomplete) list of previous 
estimates of two-photon widths see \cite{conv,GI,previous}.}! 
This gives us confidence that the approach 
provides sensible results for the two-photon production of these
mesons as well. 

\section{Form factors for two-photon production}
The differential cross section for the reaction $\re^+ \re^- \rightarrow 
\re^+ \re^- X$, where the (hadronic) final state $X$ is produced by 
$\gamma\gamma$ fusion, can (to lowest order in QED) be expressed 
in terms\footnote{We omit interference terms that integrate to zero over 
$\tilde{\phi}$, where $\tilde{\phi}$ is the azimuthal separation between
the two lepton planes in the $\gamma^\star\gamma^\star$ cms.}
of the cross sections for 
$\gamma^\star(q_1) +  \gamma^\star(q_2) \rightarrow X$ via\cite{Budnev}
\begin{equation}
  \frac{E_1\, E_2\, \rd^6\sigma}
  {\rd^3\, {\bf p}_1\, \rd^3\, {\bf p}_2} = \sum_{A,B}\, 
  {\cal L}_{A B}\, \sigma_{A B}
\ .
\end{equation}
Here $\sigma_{A B}(W,Q_1,Q_2)$ denote the cross sections 
of transverse ($A,B=T$) and longitudinal ($A,B=S$) photons
with momenta $q_i$, which depend merely on the hadronic mass 
$W = \sqrt{m_X^2}$ and the virtualities of the two photons 
$Q_i = \sqrt{-q_i^2}$. Furthermore, $p_i=(E_i,{\bf p}_i)$ are the 
momenta of the outgoing leptons, and ${\cal L}_{A B}$ are (in QED fully)
calculable virtual-photon flux factors, related to the photon-density 
matrices $\rho_i^{\lambda_1,\lambda_2}$ ($\lambda_i=\pm 1,0$). 
For example, in the standard notation\cite{Budnev}: 
\begin{equation}
 {\cal L}_{TT} = \frac{\alpha^2}{16\, \pi^4\, Q_1^2\, Q_2^2}\, 
  \frac{2\, \sqrt{X}}{s\, \sqrt{1 - 4\, m_{\re}^2/s}}\, 
  4\, \rho_1^{++}\, \rho_2^{++}
\ .
\end{equation}

Adapting the standard definition of 
the two-photon helicity cross sections 
\begin{eqnarray}
  \sigma_{TT} = & \frac{1}{4\, \sqrt{X} }\,\left[  
  W(++,++) + W(+-,+-) \right]\, , \qquad &
  \sigma_{SS} = \frac{1}{2\, \sqrt{X} }\, W(00,00)\, ,
\nonumber\\
  \sigma_{TS} = & \frac{1}{2\, \sqrt{X} }\, W(+0,+0)\, , \qquad &
  \sigma_{ST} = \frac{1}{2\, \sqrt{X} }\, W(0+,0+)\, ,
\end{eqnarray}
we find 
\begin{equation}
  W(\lambda_1,\lambda_2;\lambda_1,\lambda_2) = 
        \pi\, \delta\left( P^2 - M^2 \right) \, \left| 
      M(\lambda_1,\lambda_2) \right|^2 
\ .
\end{equation}
The amplitudes $M$ are the ones for the crossed reaction 
obtained by replacing $K_i$ by $i\, Q_i$ in the helicity amplitudes $A$. 
Note that relation (\ref{Boserel}) is changed into
\begin{equation}
  M(\lambda_1,\lambda_2) =   (-1)^{J-\lambda_1 + \lambda_2}\,
  \left.  M(\lambda_2,\lambda_1)\right|_{K_1 \leftrightarrow K_2}
\ .
\label{newBose}
\end{equation}

\begin{table}
\begin{center}
\begin{tabular}{|l|l|l|l|}
\hline
$J^P$ & $f_{TT}$ & $f_{TS}$ & $f_{SS}$
\\ \hline
$0^-$ & $\kappa\, \frac{X}{\nu^2}$ & $0$ & $0$ 
\\ \hline
$0^+$  
 & $\kappa\, \left( \frac{X + \nu\, M^2}{3\, \nu^2} \right)^2$
 & $0$
 & $2\, \kappa\, \left( \frac{M^2\, \sqrt{Q_1^2\, Q_2^2}}
                     {3\, \nu^2} \right)^2$
\\ \hline
$1^+$ 
 & $\kappa\,  \left( \frac{Q_2^2 - Q_1^2 }{2\, \nu}\right)^2$
 &  $2\, \kappa\, \frac{M^2}{2\, \nu}\,   \frac{Q_2^2}{2\, \nu}\, 
   \left( \frac{\nu + Q_1^2}{\nu} \right)^2$
 & $0$
\\ \hline
$2^+$
 &  $\kappa\, 
 \left( \frac{M^2}{2\, \nu} \right)^2\, \left\{ 1 + 
  \frac{\left[ 2\, Q_1^2\, Q_2^2 - \nu\, ( Q_1^2 + Q_2^2 ) \right]^2}
       {6\, M^4\, \nu^2 } \right\}$
 &  $ \kappa\, 
    \frac{M^2\, Q_2^2\, \left( \nu - Q_1^2 \right)^2}{4\, \nu^4}$
 &  $\kappa\,  \frac{M^4\, Q_1^2\, Q_2^2}{3\, \nu^4}$
\\ \hline
$2^-$ & $\kappa\, \left[ \frac{X}{\nu^2} \right]^3$ & $0$ & $0$ 
\\ \hline
\end{tabular}
\caption[]{Form factors $f_{AB}$.
\label{tab:FF}}
\end{center}
\end{table}
We quote the final expression for 
the cross section for the production of the $C=+1$ resonances 
in the form ($A,B=T,S$) 
\begin{equation}
 \sigma_{AB}[J^P]  = \delta\left( W^2 - M^2 \right)\, 8\, \pi^2\, 
  \frac{(2\, J+1)\,\Gamma_{\gamma\gamma}[J^P]}{M}\, 
  f_{AB}[J^P]
\ .
\label{eq:sigAB}
\end{equation}
(For the $1^+$ state we obviously use $\tilde\Gamma$.) 
The form factors $f_{AB}$ are listed in Table~\ref{tab:FF}, 
using the notation
\begin{eqnarray}
  \kappa & = & \frac{M^2}{2\, \sqrt{X}}
 \qquad \rightarrow 1 \quad {\rm for}\,\, {\rm both}
\quad Q_i^2 \rightarrow 0\, ,
\nonumber\\
  X & = & \nu^2 - Q_1^2\, Q_2^2 \, , \qquad
  \nu = \frac{1}{2}\, \left( W^2 + Q_1^2 + Q_2^2 \right) 
\ .
\end{eqnarray}

Measurements of form factors for states other than the pseudo-scalar 
mesons are still very rough \cite{Poppe,Photon97,TPC}. It is important 
to realize that the $Q_i^2$ dependence of the form factors is 
convention-dependent. What is unique is the $Q_i^2$ dependence of 
the $\re^+\re^-$ cross section, but one is free to attribute  
terms that approach $1$ for $Q_i^2 \rightarrow 0$ to either 
the luminosity functions or the 
form factors governing the two-photon cross sections. 

Conventions different from (\ref{eq:sigAB}) are in use. 
Indeed, for the pseudo-scalar mesons $P$ it has 
become standard to define the meson--photon transition form factor 
by writing the invariant amplitude as 
$M[\gamma^\star\gamma^\star \rightarrow P(0^-)] = 
F_{P\gamma\gamma}(Q_i)\, e^2\, i\, \epsilon_{\mu\nu\rho\sigma}\, 
q_1^\mu\, q_2^\nu\, \epsilon_1^\rho\, \epsilon_2^\sigma$. 
In this convention the (only non-vanishing) two-photon cross section becomes
\begin{equation}
 \sigma_{TT}[J^P]  = \delta\left( W^2 - M^2 \right)\, 8\, \pi^2\, 
  \frac{\Gamma_{\gamma\gamma}[J^P]}{M}\, \frac{1}{\kappa}\, \left[ 
         \frac{ F_{P\gamma\gamma}(Q_1,Q_2)}
              { F_{P\gamma\gamma}(0,0) } \right]^2
\ .
\label{eq:signew}
\end{equation}
Hence $F_{P\gamma\gamma}$ is related to our form factor by
\begin{eqnarray}
\frac{ F_{P\gamma\gamma}(Q_1,Q_2)}
     { F_{P\gamma\gamma}(0,0) } & = & 
  \sqrt{ \kappa\, f_{TT}[0^-] } =  \frac{M^2}{M^2 + Q_1^2  + Q_2^2}
  = \frac{M^2}{2\, \nu}\, ,
\nonumber\\
F_{P\gamma\gamma}(0,0) & = & \frac{4}{M}\, 
 \sqrt{\frac{3\, e_Q^4\, |R_{nS}(0)|^2}{\pi\, M^3}}
%  \, \frac{M^2}{M^2 + Q_1^2  + Q_2^2}
  =  \frac{2}{M}\, \sqrt{ \frac{\Gamma_{\gamma\gamma}}{\pi\, \alpha^2\, M}}
% \, \frac{M^2}{2\, \nu}
 = \frac{4\, e_Q^2\, f_P}{M^2}
\ ,
\label{OurF}
\end{eqnarray}
where $f_P$ is the pseudo-scalar decay constant, 
$f_P = |R_{S}(0)|\sqrt{\frac{3}{\pi M}}$ for heavy mesons. 
% ($f_\pi = 133\,$MeV). 
Equation (\ref{OurF}) agrees with a recent calculation \cite{Kroll97} 
where also transverse-momentum (${\bf k}_T$) effects of the quarks 
within the bound state were included. This effectively amounts to adding 
$2\, \La {\bf k}_T^2 \Ra$ to the denominator 
$M^2 + Q_1^2 + Q_2^2$ in (\ref{OurF}). 

CLEO \cite{CLEO} has measured the single $Q^2$ dependence 
($Q = Q_1$, $Q_2 \approx 0$)
of the $\pi^0$, $\eta$, and $\eta'$ form factors very precisely. 
The data are consistent with a monopole behaviour with a pole mass 
close to the $\rho$ mass for $\pi^0$ and $\eta$ and a slightly larger
mass for $\eta'$. Our result (\ref{OurF}) is indeed a monopole form factor, 
in agreement also with the power-counting rules \cite{BF}. 
We obtain good normalizations for the three pseudo-scalars, 
but the pole masses would be too small if we took the meson masses. 
On the other hand, these mesons being Goldstone bosons are 
exceptionally light. Therefore we take $m_\rho$ ($\approx$ twice  
the light quark mass) as pole mass for the three lightest pseudo-scalars, 
but identify the pole mass with the meson mass for all other mesons. 
%This is what we used for the estimate of their cross sections in 
%Table~\ref{tab:ggcross}.

It is important to realize that $F_{P\gamma\gamma}$ does not factorize  
into the product of two form factors $F_R(Q_1^2)\, F_R(Q_2^2)$ as suggested 
by vector-meson dominance. There $F_R(Q^2) = M_R^2 / (M_R^2 + Q^2)$ 
with, for example, $M_R = M_{{\rm J}/\psi}$ for charmonia. 
In particular at large $Q_1$ and $Q_2$, the form factor $F_{P\gamma\gamma}$ 
is known to fall off only as $M^2/(Q_1^2 + Q_2^2)$ \cite{BF,largeQref} 
rather than $M^4/(Q_1^2\, Q_2^2)$. 

At first sight it might seem to be a coincidence that we obtain the correct 
asymptotic behaviour, since our non-relativistic 
calculation becomes insufficient at asymptotic 
$Q_i$ values where large logarithms $\ln Q_i/M$ become important. 
Calculations with massive quarks in the non-relativistic 
approximation are well suited for $Q_i$ values not much larger than 
the heavy-quark mass $\sim M/2$. At asymptotically large $Q$, it is more 
appropriate to set up a scheme in which calculations are done
with massless quarks, but incorporating the $Q^2$ evolution of the 
quark distribution amplitudes in the meson. Such an approach is provided 
by the hard-scattering approach (HSA) \cite{BL}. 

In the HSA, the meson--photon--photon transition amplitude factorizes 
into a hard (i.e.\ perturbatively calculable) 
scattering amplitude and a soft (i.e.\ long-distance) distribution 
amplitude (DA) $\phi(x)$, so that asymptotically 
\begin{equation}
F_{P\gamma\gamma}(Q_1,Q_2) \rightarrow 2\, \La e_q^2 \Ra\, f_P\, 
  \int_0^1\, \rd x\, \frac{\phi(x)}{x\, Q_1^2 + (1-x)\, Q_2^2}
\ .
\label{asympF}
\end{equation}
Since all meson DAs approach the asymptotic form
$\phi_{\rm{as}}(x) = 6x(1-x)$, the asymptotic form of $F_{P\gamma\gamma}$ 
is fully determined in QCD
\begin{eqnarray}
 F_{P\gamma\gamma} & \rightarrow &  6\, \La e_q^2 \Ra\, f_P\, 
  \frac{Q_1^4 - Q_2^4 - 2\, Q_1^2\, Q_2^2\, \ln(Q_1^2/Q_2^2)}
       {(Q_1^2 - Q_2^2)^3}
\nonumber\\
   & \rightarrow &  \frac{6\, \La e_q^2 \Ra\, f_P}{Q^2}
  \qquad {\rm for} \,\,\ Q_1=Q,\, Q_2=0
\nonumber\\
   & \rightarrow &  \frac{2\, \La e_q^2 \Ra\, f_P}{Q^2}
  \qquad {\rm for} \,\,\ Q_1=Q_2=Q
\ .
\end{eqnarray}

Although the full $Q_i$ dependence looks more complicated than 
(\ref{OurF}), the two limiting cases show that the asymptotic 
power behaviours for large $Q_1$ and/or large $Q_2$ are identical. 
The reason is that in either calculation the hard vertex is
$\rQ\rQbar({}^1S_0) \rightarrow \gamma\gamma$ and asymptotically the 
HSA DA becomes $Q$-independent, as is the non-relativistic 
one. In fact, owing to the normalization condition 
$\int\, \rd x\, \phi(x) = 1$, the symmetric limit ($Q_1 = Q_2$) 
is fully identical. The single asymptotic limit ($Q_1 \rightarrow \infty$, 
$Q_2 = 0$) differs by a factor $2/3$, reflecting the difference of the 
moment $\La 1/x \Ra = \int\, \rd x\, \phi(x)/x$ for the asymptotic DA 
and the non-relativistic DA, $\phi_{\rm{nr}} = \delta(x - 1/2)$. 

Such numerical differences between our results and the HSA 
may also exist for the asymptotic behaviours of the other 
meson--photon transition form factors, which have not 
yet\footnote{For a recent attempt to generalize the HSA to $L\neq 0$ 
mesons, see \cite{HKPMH}.} been calculated 
in the HSA. However, we emphasize that the power fall-off is 
the same in the two approaches.  
Moreover, it is not clear at which $Q_i$ values the asymptotic regime 
is reached. Eventually, one would like to match the massive-quark 
calculation at low $Q_i$ with the HSA calculation at large $Q_i$.

\begin{table}
\begin{center}
\begin{tabular}{|c|c|c|c|c|c|c|}
\cline{2-7}
\multicolumn{1}{c}{}
 & \multicolumn{3}{|c|}{$Q_1 \neq Q_2$} 
 & \multicolumn{3}{|c|}{$Q_1 = Q_2 = Q$}
\\ \hline
$J^P$ & $F_{TT}$ & $F_{TS}$ & $F_{SS}$
 & $ F_{TT}$ & $F_{TS}$ & $ F_{SS}$
\\ \hline
$0^-$ & 1 & 0 & 0  & 1 & 0 & 0 
\\ \hline
$0^+$  
 & $\frac{\Delta}{3 }$
 & $0 $
 & $\frac{4\sqrt{2}\, M^2\, Q_1 Q_2}{3\, \Delta (Q_1^2 + Q_2^2)^2 }$
 & $\frac{2\, M}{3\, Q}$ & $0$ & $\frac{\sqrt{2}\, M}{3\, Q}$
\\ \hline
$1^+$ 
 & $1$
 & $\frac{\sqrt{2}\, M\, Q_2\, (3 Q_1^2 + Q_2^2)}{\Delta (Q_1^2 + Q_2^2)^2}$
 & $0$
 & $0$
 & $\sqrt{2}$
 & $0$
\\ \hline
$2^+$
 & $\frac{\Delta}{\sqrt{6}}$
 & $\frac{M\, Q_2 }{Q_1^2 + Q_2^2}$
 & $\frac{4\, M^2\, Q_1\, Q_2}{\sqrt{3}\, \Delta (Q_1^2 + Q_2^2)^2 }$
 & $\frac{\sqrt{42}\, M}{12\, Q}$
 & $\frac{M^2}{4\, Q^2}$
 & $\frac{M}{\sqrt{3}\, Q}$
\\ \hline
$2^-$ & $\Delta^2$ & $0$ & $0$ & $\frac{M^2}{Q^2}$ & $0$ & $0$
\\ \hline
\end{tabular}
\caption{Structure functions $F_{AB}$. 
$\Delta = |Q_2^2 - Q_1^2|/(Q_2^2 + Q_1^2)$. 
\label{DSlimit}}
\end{center}
\end{table}
The large-$Q_i$ behaviour of the form factors for the other 
$J^P$ mesons is different from that of the pseudo-scalars. 
In analogy to $F_{P\gamma\gamma}$ we define
\begin{equation}
F_{AB} = \lim_{Q_i\rightarrow \infty}\, \left[ 
  \frac{Q_1^2 + Q_2^2}{M^2}\, \sqrt{\kappa f_{AB}} \right]
\ ,
\end{equation}
and give the results in Table~\ref{DSlimit}. It can be seen that 
the double-asymptotic limit $Q_i \rightarrow \infty$ in the symmetric case 
is different from the asymmetric limit. 
For $Q_i \gg M$, but $Q_1 \neq Q_2$, the
$F_{TT}$'s are the dominant form factors; all other form factors are 
suppressed by powers of $1/\max_i Q_i$. On the contrary, $F_{SS}[0^+]$ 
possesses the same power counting $\propto 1/Q$ as $F_{TT}[0^+]$ 
in the symmetric case. The same holds for $2^+$ mesons, while 
$F_{TT}[2^-]$ behaves as $1/Q^2$ and Bose symmetry leads 
to a vanishing of $F_{TT}[1^+]$. In fact, besides $F_{TT}[0^-]$ 
only $F_{TS}[1^+]$ remains non-zero in the symmetric 
double-asymptotic limit. Therefore the cross sections for
$1^+$ mesons, which vanish at zero $Q_i$, become the largest $P$-wave 
cross sections at high $Q_i$ in the symmetric limit.
 
Ideally the aim is to measure the dependence on both $Q_1$ and $Q_2$, 
and to separate the form factors $f_{SS}$, $f_{TS}$, and $f_{TT}$ 
(as well as the latter in helicity-two and 
helicity-zero components). Obviously, such tasks require high statistics and 
excellent tagging efficiencies. Experimentally much more feasible are  
single-tag measurements, $\rd \sigma/\rd Q_2^2$. Often an 
anti-tag is imposed on the other electron in order to ensure
$Q_1 \approx 0$. Such measurements are sensitive to only 
an effective form factor. We therefore generalize the pseudo-scalar 
form factor $F_{P\gamma\gamma}$ and define
\begin{equation}
  \left( F_{\rm eff}(Q^2) \right)^2  =  \left. \lim_{Q_1^2 \rightarrow 0}\, 
\kappa [  f_{TT} + \epsilon\,  f_{TS}] \right|_{Q_2^2=Q^2}
\ ,
\end{equation}
where $\epsilon$ is the ratio of the average luminosities 
${\cal L}_{TS}$ to ${\cal L}_{TT}$. Experimentally, $\epsilon$ is
often close to $1$. We find:
\begin{eqnarray}
    F_{\rm eff}[0^-] & = & \frac{M^2}{M^2 + Q^2}
\nonumber\\
    F_{\rm eff}[0^+] & = & \frac{1}{3}\, \frac{M^2}{M^2 + Q^2} \, 
   \left( 1 + \frac{2\, M^2}{M^2 + Q^2} \right)
 \stackrel{Q \rightarrow 0}{\rightarrow} \frac{M^2}{M^2 + Q^2}\, 
  \left[ 1 - \frac{2}{3}\, \frac{Q^2}{M^2} \right]
\nonumber\\
    F_{\rm eff}[1^+] & = & \sqrt{2}\, \left(\frac{M^2}{M^2 + Q^2}\right)^2\, 
  \left[  \frac{Q^2}{M^2}\,  
  \left\{ \epsilon +  \frac{Q^2}{2\, M^2} \right\} \right]^{1/2}
 \stackrel{Q \rightarrow 0}{\rightarrow} \frac{M^2}{M^2 + Q^2}\, 
  \sqrt{2\, \epsilon}\, \frac{Q}{M} 
\nonumber\\
    F_{\rm eff}[2^+] & = & \frac{M^2}{M^2 + Q^2}\, 
  \left[ \frac{ Q^4 + 6\, M^4 + 6\, \epsilon\, M^2\, Q^2}{6\, (M^2+Q^2)^2}
  \right]^{1/2}
 \stackrel{Q \rightarrow 0}{\rightarrow} \frac{M^2}{M^2 + Q^2}\, 
  \left[ 1 - \frac{2-\epsilon}{2}\, \frac{Q^2}{M^2} \right]
\nonumber\\
    F_{\rm eff}[2^-] & = & \frac{M^2}{M^2 + Q^2}
\ .
\label{Feffdef}
\end{eqnarray}
Two features are evident. First, for $Q$ small compared to $M$, all
but the $1^+$ form factors are similar to the one of pseudo-scalar mesons. 
Secondly, all form factors asymptotically behave as $1/Q^2$, but 
the hierarchy changes, $Q^2\, F_{\rm eff}/M^2 \rightarrow 
1$, $1/3$, $1$, $1/\sqrt{6}$, $1$. Hence, at large $Q^2$ we predict
\begin{equation}
\frac{M\, \sigma[\re^+\re^-]}{\Gamma_{\gamma\gamma}} 
= 1: \frac{1}{9} : 3 : \frac{5}{6} : 5
\quad {\rm  for} \quad J^P = 0^- : 0^+ : 1^+ : 2^+ : 2^- 
\label{largeQratio}
\ .
\end{equation}

We close this section by comparing our results with previous 
calculations. We have already commented upon the pseudo-scalar case. 
Our covariant $P$-wave amplitudes agree with 
the $J=1$ amplitude of \cite{Cahn}, the $J=2$ amplitude 
of \cite{KV}, the amplitudes of \cite{KKS} for all $J$, but disagree
with the $J=0$ expression of \cite{KV}. For $k_i^2 =0$ we reproduce 
the $D$-wave result of \cite{Cho}. Concerning the form factors, 
we are not aware of calculations of the $0^{+}$ and $2^-$ form factors. 
%{\bf this is serious: can you ask an experimental colleague?} 
An effective form factor for the $2^+$ state is given in \cite{KV} 
without quoting the value of $\epsilon$. We reproduce their result 
for (the unusual value) $\epsilon = -1/2$. 
The $1^+$ form factors were calculated in \cite{Cahn}. We find 
identical intermediate results, (4.4) and (4.5) in \cite{Cahn},  
but the final form factors were given in the single-tag limit only, 
multiplied by ad hoc vector-meson-dominance form factors 
($F_\rho(Q^2) = m_\rho^2/(m_\rho^2 + Q^2)$); for example
\begin{equation}
  f_{TS} = 2\, \frac{2\, \sqrt{X}}{M^2}\, \frac{Q_2^2}{M^2}\, 
  F^2(Q_1^2)\,  F^2(Q_2^2)
\ .
\end{equation}
This result was then used by TPC/$2\gamma$ 
to define a form factor $\tilde{F}$ for single-tag $1^+$ events
\begin{equation}
 \tilde{F}^2 = F_{\rho}^2(Q_1^2)\, F_{\rho}^2(Q_2^2)\, \frac{Q_2^2}{M^2}\, 
  \left\{ 1 + \epsilon^{-1}\, \frac{Q_2^2}{2\, M^2} \right\} 
\ .
\label{eqTPC}
\end{equation}
Our result for this form factor is
\begin{eqnarray}
  \tilde{F}^2 & \equiv & \frac{\epsilon}{2}\, 
  \kappa \, \left( f_{ST} + \epsilon^{-1}\, 
  f_{TT} \right)
\nonumber\\
  & = & \left(\frac{M^2}{2\, \nu} \right)^4\, \left\{ 
  \frac{Q_2^2}{M^2}\, \left( \frac{\nu+Q_1^2}{\sqrt{X}} \right)^2 
  + \epsilon^{-1}\, \frac{ (Q_1^2-Q_2^2)^2}{2\, M^4}\, 
  \frac{\nu^2}{X} \right\}
\nonumber\\
  & \stackrel{Q_1 \ll Q_2}{\rightarrow} & 
\left(\frac{M^2}{2\, \nu}\right)^4\, \frac{Q_2^2}{M^2}\, 
  \left\{ 1 + \epsilon^{-1}\, \frac{Q_2^2}{2\, M^2} \right\}
\ .
\end{eqnarray}
The $(M^2/\nu)^4$ factor is in fact crucial to obtain a sensible 
behaviour at large $Q_2^2$, in contrast to (\ref{eqTPC}), which approaches 
a constant as $Q_2 \rightarrow \infty$.
%Hence we would agree if we replaced $\kappa^4$ by 
%$F^2_\rho(Q_1^2)\, F^2_\rho(Q_2^2)$. 

%%%%%%%%%%%%%%%%%%%%%%%%%%%%%%%%%%%%%%%%%%%%%%%%%%%%%%%%%%%%%%%%%%%%
\section{Cross-section results}
\begin{table}
\begin{center}
  \begin{tabular}{|c|c|c|c|c|c||c|} \hline
$i \backslash j$& $I=1$ & $ I=0       $ & $ (I=0)'   $ & $ \rc \bar{\rc} $ 
& $ \rb \bar{\rb}$ & $\rc \bar{\rc}$ [$Q_2^2<0.1\,$GeV$^2$] \\ \hline
$0^-(1S) $ & $2.9$ & $1.7$ & $3.9$ & $0.12$ & $0.13/10^3$ & $0.10$  \\ \hline
$0^+ $ & $2.9$ & $0.42$ & $2.0$ & $0.024$ & $0.99/10^5$ & $0.020$\\ \hline
$1^+ $ & $0.060$ & $0.14$ & $0.35/10^2$ & $0.96/10^3$ & $0.43/10^6$
  & $0.38/10^3$\\ \hline
$2^+ $ & $1.1$ & $3.7$ & $0.085$ & $0.012$ & $0.85/10^5$
  & $0.98/10^2$\\ \hline
$2^- $ & $0.69$ & $0.23$ & $1.8$ & $0.31/10^3$ & $0.81/10^7$
  & $0.26/10^3$\\ \hline
$0^-(2S) $ & $1.7$ & $0.56$ & $4.4$ & $0.029$ & $0.47/10^4$
  & $0.024$\\ \hline
  \end{tabular}
\end{center}
\caption{Cross sections for $\re^+\re^- \rightarrow R[J^P]$ in nb 
for $E_b = 91.25\,$GeV; no-tag columns 2--5, single anti-tag last column.
\label{tab:ggcross}}
\end{table}
Numerical results for a representative LEP energy are given in
Table~\ref{tab:ggcross}, appropriate for a no-tag setup, i.e.\ 
no cuts have been applied to the scattered leptons. Most event rates 
are sizeable, although experimental acceptance and tagging efficiency 
will somewhat lower the total rates \cite{LEP2}. 
In particular we find 
rates of about $1\,$pb for the $\chi_{\rc 1}$, $0.1\,$pb for the 
$\eta_{\rb}$, and $0.03\,$nb for the $\eta_{\rc}(2S)$. 

\begin{figure}
\begin{center}
%\begin{tabular}{cc}
%\epsfig{figure=job_q2_chic.eps,width=0.48\textwidth}
\epsfig{figure=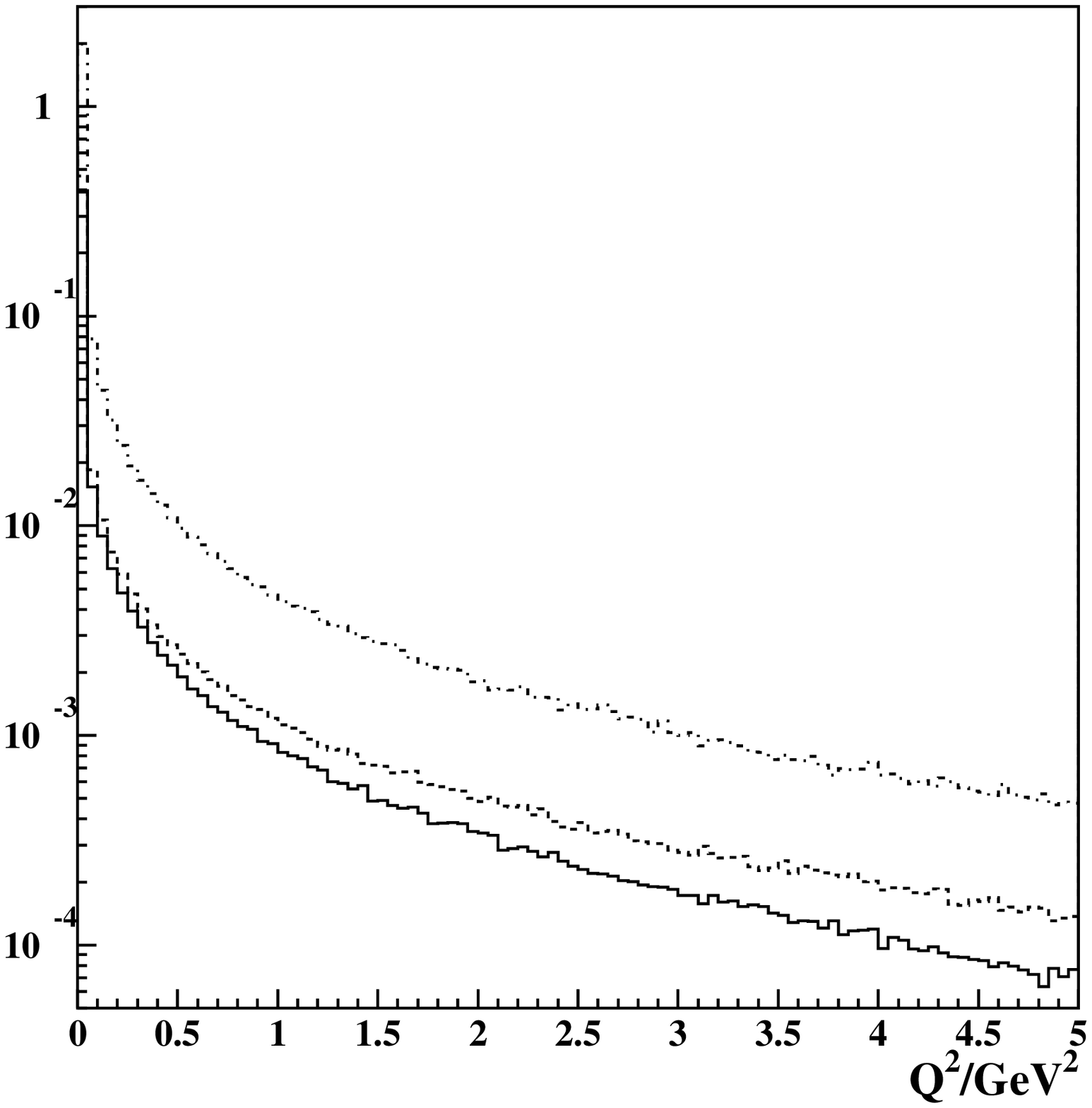,width=0.44\textheight}
\epsfig{figure=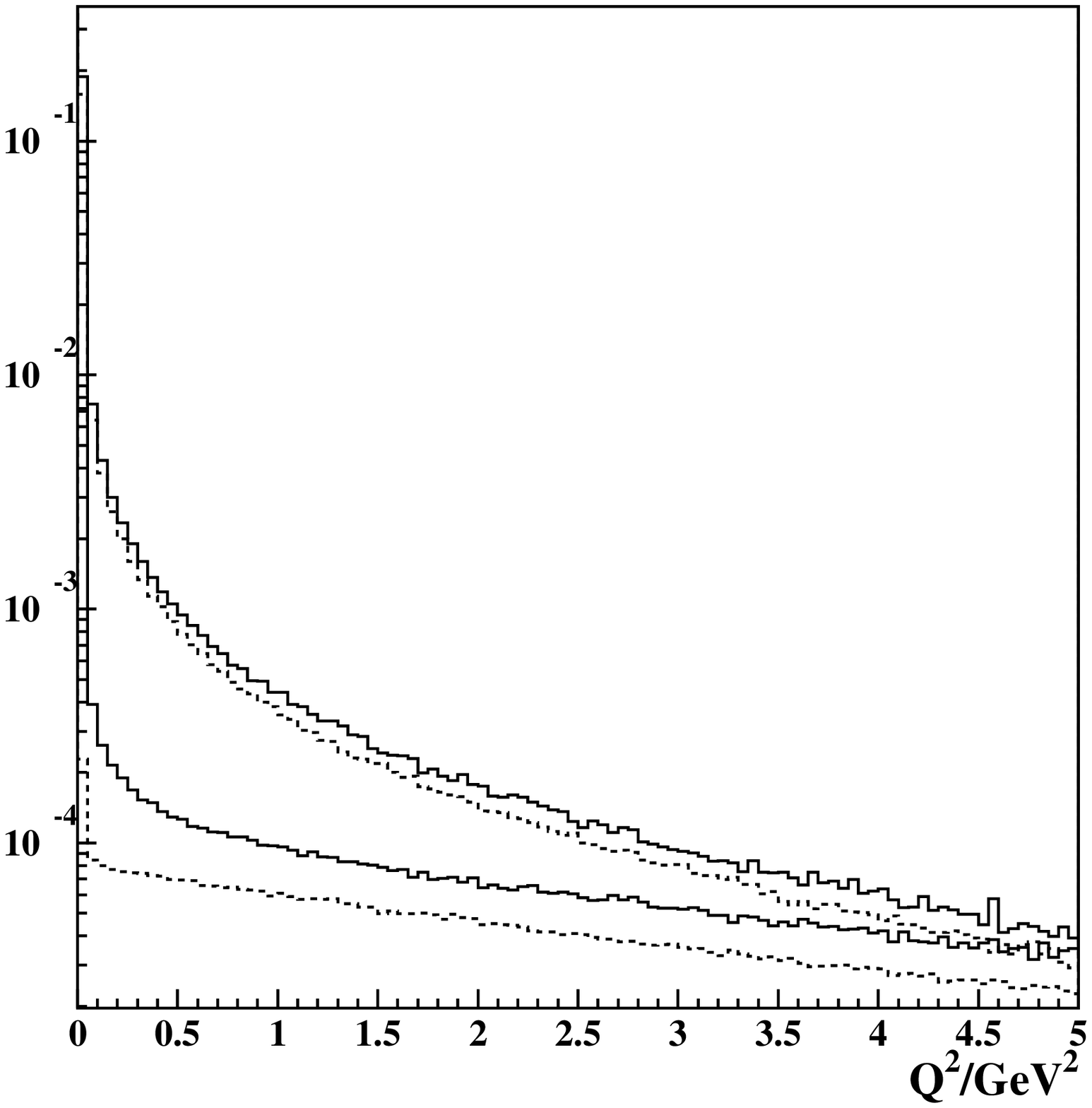,width=0.44\textheight}
% &
%\epsfig{figure=job_lnq2_chic.eps,width=0.45\textheight} 
%\end{tabular}
\caption[]{Top: Cross section $\rd \sigma/ \rd Q_1^2 
[\re^+\re^- \rightarrow \re^+\re^-\, M]$ in nb/GeV 
at $\sqrt{s} = 92.5\,$GeV 
for $\eta_{\rc}(1S)$ (dash-dotted), 
$\eta_{\rc}(2S)$ (dashed), 
and $\chi_{\rc0}$ (solid). 
%\label{fig:etac}}
%\end{center}
%\end{figure}
%%%%%%%%%%%%%%%%%%%%%%%%%%%%%%%%%%%%%%%%%%%%%%%%%%%%%%%
%\begin{figure}
%\begin{center}
%%\begin{tabular}{cc}
%%\epsfig{figure=job_q2_chic.eps,width=0.48\textwidth}
%\epsfig{figure=job_q2_chic.eps,width=0.45\textheight}
%% &
%%\epsfig{figure=job_lnq2_chic.eps,width=0.45\textheight} 
%%\end{tabular}
%\caption[]{Cross section $\rd \sigma/ \rd Q_1^2 
%[\re^+\re^- \rightarrow \re^+\re^-\, \chi_{\rc J}]$ 
%at $\sqrt{s} = 92.5\,$GeV 
%for $J=2$ ($J=1$) without $Q_2^2$ cut as solid (dashed) line, 
%for $J=2$ ($J=1$) with $Q_2^2 < 0.1\,$GeV$^2$ as dotted (dash-dotted) line. 
%Bottom: Same but $(1/Q^2)\, \rd \sigma /\rd \log_{10}(Q^2)$.
Bottom: Ditto but for $\chi_{\rc 2}$ (upper lines) and $\chi_{\rc 1}$ 
(lower lines); dashed lines for $Q_2^2 < 0.1\,$GeV$^2$.
\label{fig:diff}}
\end{center}
\end{figure}

\begin{figure}
\begin{center}
\epsfig{figure=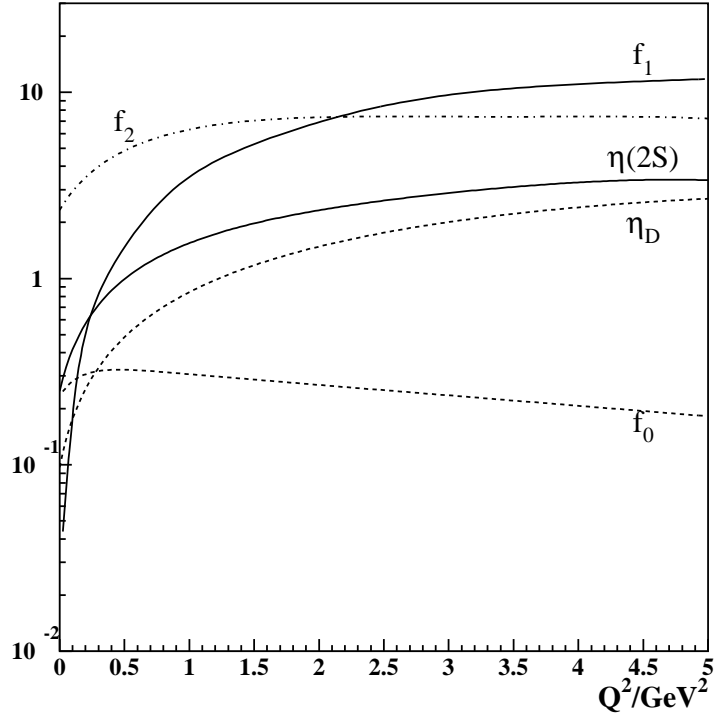,width=0.44\textheight} 
\epsfig{figure=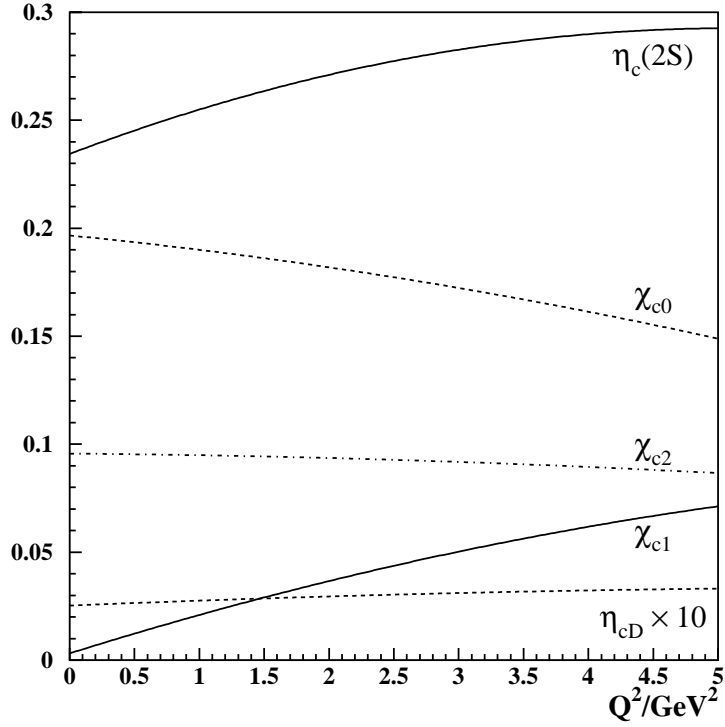,width=0.44\textheight} 
\caption[]{Cross-section ratios 
$\rd \sigma[J^P] / \rd Q_1^2$ over $\rd \sigma[0^-(1S)] / \rd Q_1^2 $
in $\re^+\re^- \rightarrow \re^+\re^-\, M(J^P)$ 
at $\sqrt{s} = 92.5\,$GeV 
for the $\eta$ family (top) and the $\eta_{\rc}$ family (bottom). 
%$0^-(2S)$ (solid), $0^+$ (dashed), $1^+$ (solid), $2^+$ (dash-dotted), 
%$2^-$ (dashed). 
The $\eta_{\rc D}/\eta_{\rc}(1S)$ ratio is multiplied by $10$.
\label{fig:rat6}}
\end{center}
\end{figure}

The single-$Q^2$ distribution of charmonium production is shown in 
Fig.~\ref{fig:diff}, $\rd  \sigma/ \rd Q_1^2$. Here we have 
integrated over all $Q_2$. All distributions are steeply 
falling with $Q_1^2$, even for the $\chi_{\rc 1}$. 
Hence the modulation of the $Q_i$ dependence of the luminosity functions 
(mainly due to the $1/Q_i^2$ photon propagator) by the form factors 
is rather weak. This is more clearly seen in Fig.~\ref{fig:rat6}, where
we show ratios of $\re^+\re^-$ cross sections as functions of 
$Q_1^2$, taking as  reference the $\eta_{\rc}$ distribution. 
This behaviour was anticipated in (\ref{Feffdef}): for $Q \ll M$, 
the effective single-tag form factors all resemble that of the 
pseudo-scalar meson. 

The situation is quite different for $Q \gg M$, accessible through 
studies of light mesons (Fig.~\ref{fig:rat6}). In fact, the cross-section 
ratios approach the asymptotic values (\ref{largeQratio}) already at 
$Q_1^2 \approx 5\,$GeV$^2$, for example, 
$\rd \sigma[\rf_1]/ \rd \sigma[\eta] \sim 10$ and 
$\rd \sigma[\rf_0]/ \rd \sigma[\eta] \sim 0.1$. 
%Even $\rf_1/\eta$ and $\chi_{\rc 1}/\eta_{\rc}$ not steep, as 
%clear from Fig.~\ref{fig:diff}. 

\begin{figure}
\begin{center}
\epsfig{figure=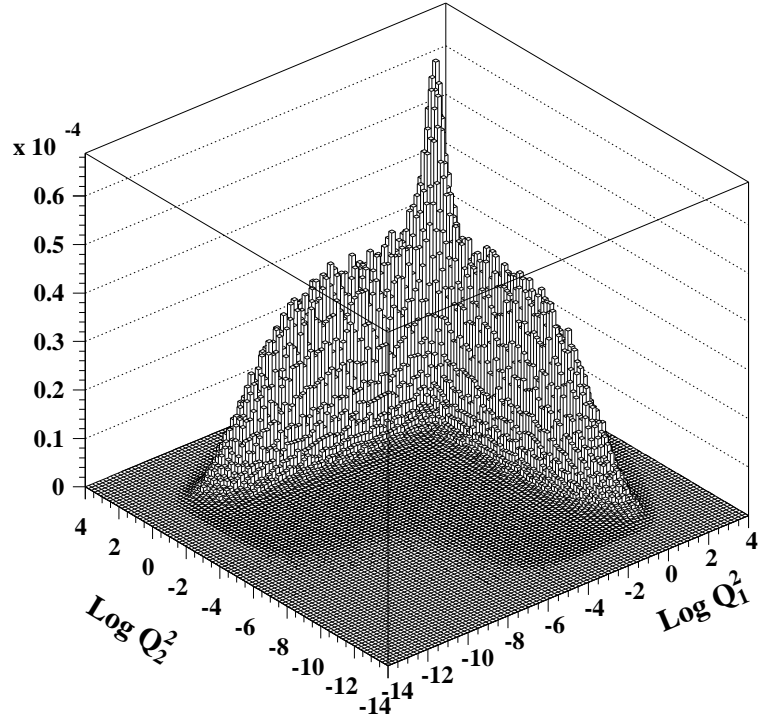,width=0.45\textheight} 
\epsfig{figure=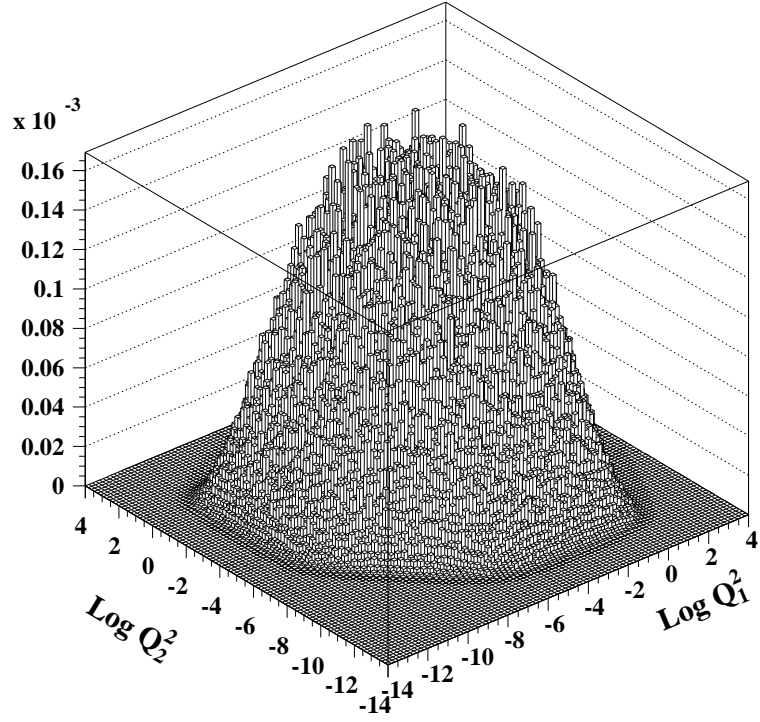,width=0.45\textheight} 
\caption[]{Cross-section  
$\rd^2 \sigma / \rd \log_{10}(Q_1^2/ {\rm GeV}^2) \, 
\rd \log_{10}(Q_2^2/ {\rm GeV}^2) $ in nb
of $\re^+\re^- \rightarrow \re^+\re^-\, \chi_{\rc J}$ 
at $\sqrt{s} = 92.5\,$GeV 
for $J=1$ (top) and $J=2$ (bottom).
\label{fig:rat5}}
\end{center}
\end{figure}

In Fig.~\ref{fig:rat5} we compare $\chi_{\rc 1}$ production with 
$\chi_{\rc 2}$ production. Tensor mesons are dominated by phase-space 
regions where both photon virtualities $Q_i$ are small. On the other 
hand, this region is suppressed for axial-vector mesons. 
However, this difference is pronounced only at very low $Q^2$, visible 
on logarithmic scales 
$\rd^2\, \sigma/ \rd \log_{10}Q_1^2\, \rd \log_{10}Q_2^2$ 
(Fig.~\ref{fig:rat5}).   
For example, $Q_1^2 = 10^{-1}\,$GeV$^2$ is large enough to reach very 
small $Q_2$ values. Hence the single-$Q^2$ distribution is still peaked 
at low $Q^2$. 

\begin{figure}
\begin{center}
\epsfig{figure=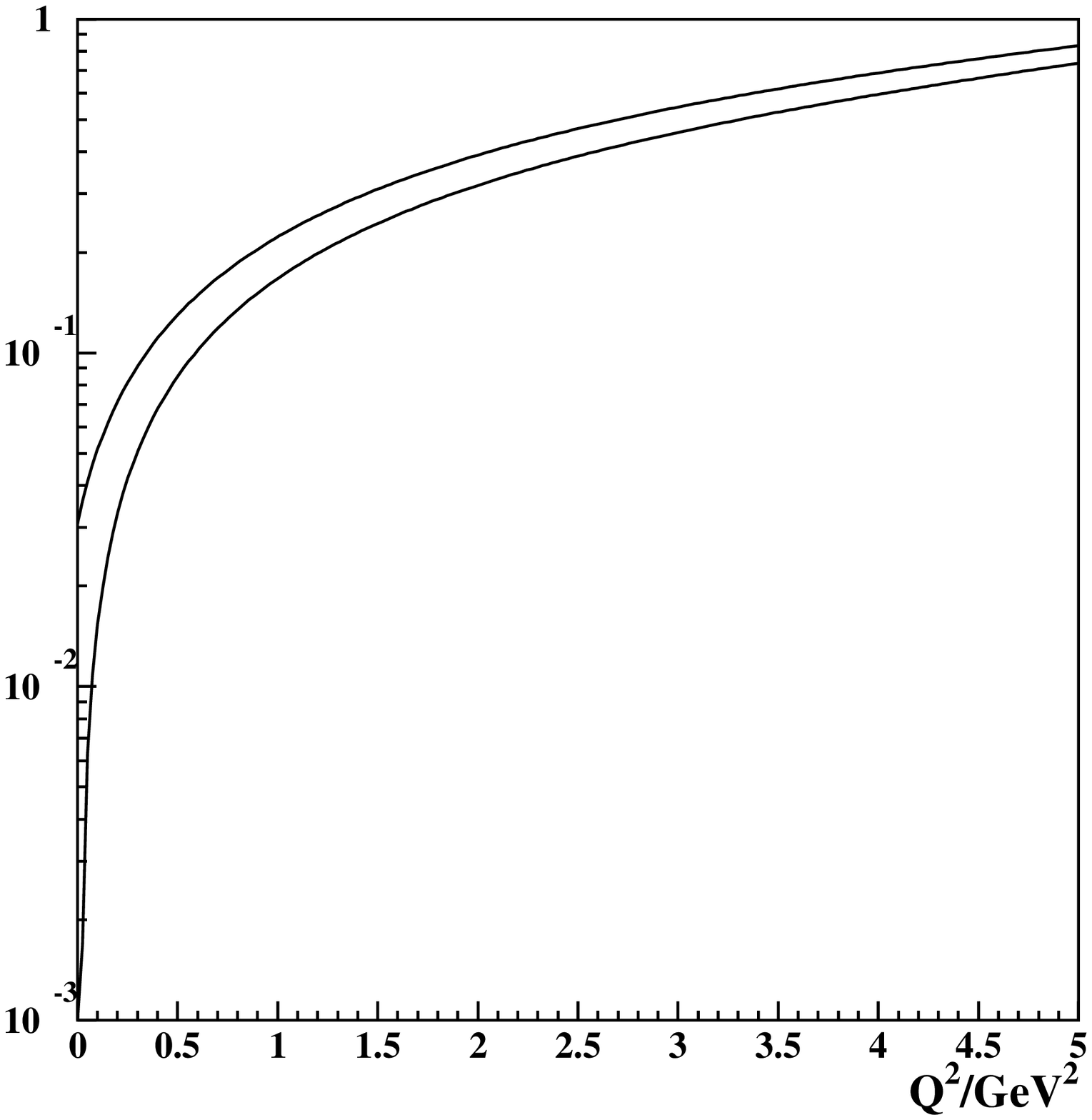,width=0.44\textheight} 
\epsfig{figure=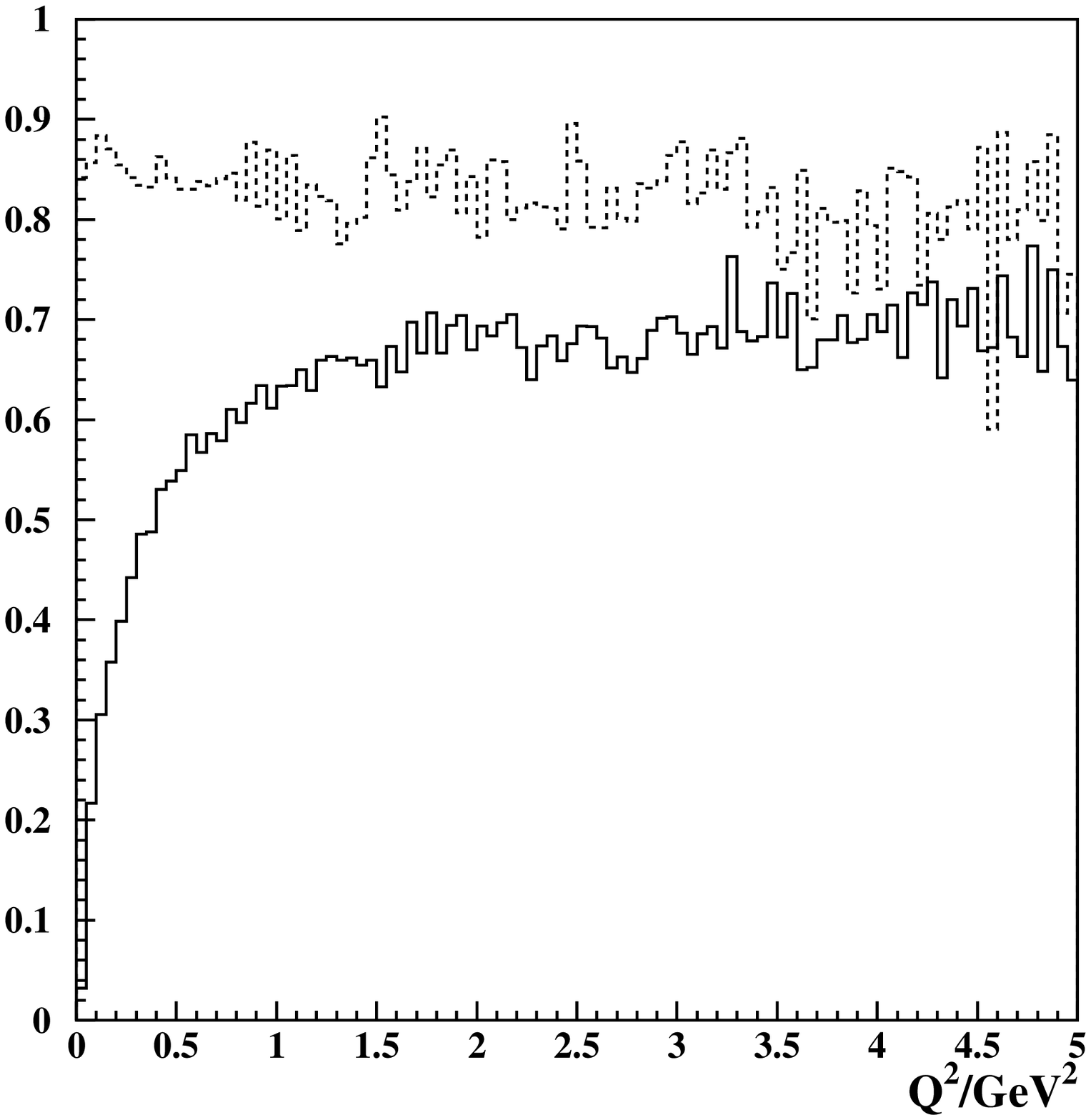,width=0.44\textheight}
\caption[]{Cross-section ratios
in $\re^+\re^- \rightarrow \re^+\re^-\, \chi_{\rc J}$ 
at $\sqrt{s} = 92.5\,$GeV. Top:  
$\chi_{\rc 1}/\chi_{\rc 2}$ with $Q_2^2 < 0.1\,$GeV$^2$ cut 
(lower line) and without $Q_2^2$ cut (upper line).
%(solid) and without $Q_2^2$ cut (dashed).
%Bottom: Same but on a logarithmic scale.
%\label{fig:rat1}}
%\end{center}
%\end{figure}
%%%%%%%%%%%%%%%%%%%%%%%%%%%%%%%%%%%%%%%%%%%%%%%%%%%%%%
%\begin{figure}
%\begin{center}
%\epsfig{figure=job_rat3_chic.eps,width=0.45\textheight}
%\caption[]{Cross-section ratio $\chi_{\rc 1}/\chi_{\rc 2}$ 
%in $\re^+\re^- \rightarrow \re^+\re^-\, \chi_{\rc J}$ 
%at $\sqrt{s} = 92.5\,$GeV 
%with $Q_2^2 < 0.1\,$GeV$^2$ (solid) and 
%without $Q_2^2$ cut (dashed) as a function of $\log_{10}(Q^2)$. 
%\label{fig:rat3}}
%\end{center}
%\end{figure}
%
%\begin{figure}
%\begin{center}
%\epsfig{figure=job_rat4_chic.eps,width=0.45\textheight}
%\epsfig{figure=job_rat5_chic.eps,width=0.45\textheight} 
%\caption[]{Cross-section ratio $R(Q_1^2) = 
%\rd \sigma / \rd Q_1^2[\chi_{\rc J}(Q_2^2 < 0.1) ] / 
%\rd \sigma / \rd Q_1^2[\chi_{\rc J}(Q_2^2 < s  ) ]$
%in $\re^+\re^- \rightarrow \re^+\re^-\, \chi_{\rc J}$ 
%at $\sqrt{s} = 92.5\,$GeV 
%for $J=1$ (solid) and $J=2$ (dashed).
%Bottom: Same as a function of $\log_{10}(Q^2)$. 
Bottom: $\chi_{\rc 2}\, [Q_2^2 < 0.1\,{\rm GeV}^2] 
/ \chi_{\rc 2}\, [{\rm no}\,\, {\rm cut}]$ (dashed) and 
$\chi_{\rc 1}\, [Q_2^2 < 0.1\,{\rm GeV}^2] 
/ \chi_{\rc 1}\, [{\rm no}\,\, {\rm cut}]$ (solid).
\label{fig:rat4}}
\end{center}
\end{figure}

In particular it follows that an anti-tag cut, say $Q_2^2 < 0.1\,$GeV$^2$, 
is no means to enrich $1^+$ states in $2\gamma$ event samples. 
In fact, the upper $Q_2$ cut has a stronger effect 
on the $1^+$ than on the $2^+$ mesons, see Fig.~\ref{fig:rat4}. 
This can be seen also in Table~\ref{tab:ggcross}, where 
in the last column we quote charmonium rates after 
applying an upper limit of $0.1\,$GeV$^2$ on $Q_2^2$. 
%It is important 
%to realize that such a cut reduces the $1^+$ rate much more than 
%the others. It is only at large $Q_1^2$ that the ratio of 
%$\sigma[1^+]$ to, say, $\sigma[2^+]$ is enhanced.
Hence it appears sensible to search for $1^+$ mesons also 
in no-tag event samples and/or single-tag samples where 
no anti-tag is applied to the other electron.

Since we feel that the $Q^2$ behaviour of $1^+$ mesons is not yet 
fully appreciated, we want to exhibit its $Q^2$ dependence also analytically. 
For studies of the low-$Q_i$ behaviour of $2\gamma$ processes in 
$\re^+\re^-$ collisions it is convenient to consider 
the approximation of keeping only the leading $Q_i$ dependences 
(Weizs\"acker--Williams or equivalent-photon approximation). 
Then the luminosity functions simplify to\footnote{
We stress that for our numerical results we keep the full dependence 
on $Q_i$, in which case the ${\cal L}_{AB}$ do not factor into products 
of $Q_1$-dependent and $Q_2$-dependent functions.} 
\begin{equation}
 {\cal L}_{A,B}\, \left( \prod_i \, \frac{ \rd^3 {\bf p}_i}{E_i} \right)
  = \left( \frac{\alpha}{2\, \pi}\right)^2\, \frac{2\, \sqrt{X}}{W^2}\, 
  \frac{\rd Q_1^2}{Q_1^2}\,  \frac{\rd Q_2^2}{Q_2^2}\,  
  \frac{\rd x_1}{x_1}\,  \frac{\rd x_2}{x_2}\,
  Y_A(x_1)\,   Y_B(x_2)
\ ,
\label{WWA}
\end{equation}
where
\begin{equation}
  Y_T(x) = \frac{2\, (1-x)(1 - Q_{\rm{min}}^2/Q^2 ) + x^2}{x}\ , 
\qquad Y_S(x) = \frac{2\, (1-x)}{x}
\ .
\end{equation}
Hence the $\re^+\re^-$ cross section must not vanish at low $Q_i$
even if the two-photon cross section vanishes for $Q_i \rightarrow 0$: 
multiplying (\ref{WWA}) by $\sigma_{TS}$ and $\sigma_{ST}$, the dominant 
behaviour at low $Q_i$ of the $1^+$ cross section is
\begin{equation}
  \sigma_{\re\re} = 
 \int\, \rd Q_1^2\, \int\, \frac{\rd Q_2^2}{Q_2^2}\, c_{ST} + 
 \int\, \frac{\rd Q_1^2}{Q_1^2}\, \int\, \rd Q_2^2\, c_{TS} + \ldots
\end{equation}
In fact, $\rd \sigma_{\re\re}[1^+]/\rd Q_1$ is even peaked at low $Q_1$. 

%{\it Acknowledgements:}\hfill\\
%{\bf Do we want/have to thank anyone?}

%

\clearpage
%%%%%%%%%%%%%%%%%%%%%%%%%%%%%%%%%%%%%%%%%%%%%%%%%%%%%%%%%%%%%
%%%%%%%%%%%%%%%%%%%%%%%%%%%%%%%%%%%%%%%%%%%%%%%%%%%%%%%%%%%%%
%%%%%%%%%%%%%%%%%%%%%%%%%%%%%%%%%%%%%%%%%%%%%%%%%%%%%%


\begin{thebibliography}{99}
%
\bibitem{BL}
S.J.\ Brodsky and G.P.\ Lepage, Phys.\ Rev.\ {\bf D22} (1980) 2157; 
ibid.\ {\bf D24} (1981) 1808.
\bibitem{ref:PionFF}
S.J.\ Brodsky, C.-R.\ Ji, A.\ Pang and D.G.\ Robertson, 
SLAC-PUB-7473, May 1997, hep-ph/9705221; 
%\hfill\\{\it OPTIMAL RENORMALIZATION SCALE AND 
%SCHEME FOR EXCLUSIVE PROCESSES}
%
\hfill\\
P.\ Kroll and M.\ Raulfs, Phys.\ Lett.\ {\bf B387} (1996) 848;
%\hfill\\{\it THE PI GAMMA TRANSITION FORM-FACTOR AND THE PION WAVE FUNCTION}
%
\hfill\\
S.\ Ong, Phys.\ Rev.\ {\bf D52} (1995) 3111;
%\hfill\\{\it IMPROVED PERTURBATIVE QCD ANALYSIS OF 
%THE PION - PHOTON TRANSITION FORM-FACTOR}
%
\hfill\\
A.V.\ Radyushkin and R.\ Ruskov,  JLAB-THY-97-24, June 1997, 
hep-ph/9706518; 
\hfill\\
%{\it  THE ASYMPTOTICS OF THE TRANSITION FORM-FACTOR 
%GAMMA GAMMA* ---> PI0 AND QCD SUM RULES}
%
A.\ Anselm et al., HUTP-95/A037, November 1995, hep-ph/9603444.
%
\bibitem{Kroll97}
Th.\ Feldmann and P.\ Kroll,  Univ.\ of Wuppertal preprint 
WUB-97-18, May 1997, hep-ph/9706224. 
%\hfill\\{\it PREDICTIONS FOR THE ETA(C) GAMMA TRANSITION FORM-FACTOR}
\bibitem{Reviews}
For recent reviews see, for example, 
S.J.\ Brodsky, SLAC-PUB-7604, July 1997, hep-ph/9708345; 
%\hfill\\{\it EXCLUSIVE PHOTON-PHOTON PROCESSES}
%
\hfill\\
G.\ Sterman and P.\ Stoler, ITP-SB-97-49, August 1997, hep-ph/9708370. 
%\hfill\\{\it HADRONIC FORM-FACTORS AND PERTURBATIVE QCD}
%
\bibitem{KWZ}
G.\ K\"opp, T.F.\ Walsh and P.\ Zerwas, 
Nucl.\ Phys.\ {\bf B70} (1974) 461
%
\bibitem{KV}
H.\ Krasemann and J.A.M.\ Vermaseren, Nucl.\ Phys.\ {\bf B184} (1981) 269.
%
\bibitem{Cahn}
R.N.\ Cahn, Phys.\ Rev.\ {\bf D35} (1987) 3342; ibid.\ {\bf D37} (1988) 833.
%
\bibitem{Poppe}
M.\ Poppe, J.\ Mod.\ Phys.\ {\bf A1} (1986) 545. 
\bibitem{Photon97}
D.\ Morgan et al., J.\ Phys.\ G: Nucl.\ Part.\ Phys.\ {\bf 20} (1994) A1.
%Proc.\ of the workshops on $\gamma\gamma$ collisions, 
%Photon '95, Sheffield 1995, World Scientific, eds.\ D.J.\ Miller et al.,
%
\bibitem{GALUGA}
G.A.\ Schuler, CERN-TH/96-313, hep-ph/9611249, 
Comput.\ Phys.\ Commun.\ in press.
%
\bibitem{BBL}
G.T.\ Bodwin, E.\ Braaten and G.P.\ Lepage, Phys.\ Rev.\ {\bf D51} 
(1995) 1125. 
%
\bibitem{KKS}
J.H.\ K\"uhn, J.\ Kaplan and E.G.O.\ Safiani, 
Nucl.\ Phys.\ {\bf B157} (1979) 125. 
%
\bibitem{Novikov}
V.A.\ Novikov et al., Phys.\ Rep.\ {\bf 41} (1978) 1.
%
\bibitem{TPC}
TPC/$2\gamma$ collab., H.\ Aihara et al., Phys.\ Rev.\ {\bf D38} (1988) 1.
%
\bibitem{EQ}
E.J.\ Eichten and C.\ Quigg, Phys.\ Rev.\ {\bf D52} (1995) 1726.
%
\bibitem{PDG}
Review of Particle Physics, Phys.\ Rev.\ {\bf D54} (1996) 1.
%
\bibitem{conv}
E.S.\ Ackleh and T.\ Barnes, Phys.\ Rev.\ {\bf D45} (1992) 232;\hfill\\
C.\ Hayne and N.\ Isgur, Phys.\ Rev.\ {\bf D25} (1982) 1944. 
%
\bibitem{GI}
S.\ Godfrey and N.\ Isgur, Phys.\ Rev.\ {\bf D32} (1985) 189.
%
\bibitem{previous}
C.R.\ M\"{u}nz, Nucl.\ Phys.\ {\bf A609} (1996) 364;\hfill\\
J.D.\ Anderson et al., Phys.\ Rev.\ {\bf D43} (1991) 2094.
%
\bibitem{Budnev}
V.M.\ Budnev et al., Phys.\ Rep.\ {\bf 15} (1975) 181.
%
\bibitem{CLEO}
CLEO collab., J.\ Gronberg et al., preprint CLNS 97/1477, July 1997.
%
\bibitem{BF}
S.J.\ Brodsky and G.R.\ Farrar, Phys.\ Rev.\ {\bf D11} (1975) 1309.
%
\bibitem{largeQref}
V.A.\ Novikov et al., Nucl.\ Phys.\ {\bf B237} (1984) 525;\hfill\\
P.\ Kessler and S.\ Ong, Phys.\ Rev.\ {\bf D48} (1993) 2974.
%
\bibitem{HKPMH}
L.\  Houra-Yaou, P.\ Kessler, J.\ Parisi, F.\ Murgia and J.\ Hansson, 
LPC-96-53, November 1996, hep-ph/9611337. 
%\hfill\\{\it PRODUCTION OF MESON PAIRS, INVOLVING TENSOR AND 
%PSEUDOTENSOR MESONS, IN PHOTON-PHOTON COLLISIONS}
%
\bibitem{Cho}
P.\ Cho and M.B.\ Wise, Phys.\ Rev.\ {\bf D51} (1995) 3352.
%
\bibitem{LEP2}
P.\ Aurenche, G.A.\ Schuler (Conveners), $\gamma\gamma$ Physics, 
in Proc.\ Workshop on Physics at LEP2, eds.\ G.\ Altarelli et al.\ 
(CERN 96-01, Geneva, 1996), 
Vol.~1, p.~291.
\end{thebibliography}
\end{document}